# Improving Malaria Parasite Detection from Red Blood Cell using Deep Convolutional Neural Networks

Aimon Rahman[1], Hasib Zunair[1], M Sohel Rahman[2*], Jesia Quader Yuki[1], Sabyasachi Biswas[1], Md Ashraful Alam[3], Nabila Binte Alam[4], M.R.C. Mahdy[1]*

[1] *Department of Electrical & Computer Engineering, North South University, Bashundhara, Dhaka, 1229, Bangladesh.*

[2]*Department of Computer Science & Engineering, Bangladesh University of Engineering and Technology ECE Building, West Palasi, Dhaka-1205*

[3]*Department of Pharmaceutical Sciences, North South University, Bashundhara, Dhaka, 1229, Bangladesh*

[4] *Department of Computer Science & Engineering, Military Institute of Science and Technology, Mirpur 12, Dhaka, Bangladesh*

*Corresponding Authors: mahdy.chowdhury@northsouth.edu and msrahman@cse.Buet.ac.bd

**Malaria is a female anopheles' mosquito-bite inflicted life-threatening disease which is considered endemic in many parts of the world. This article focuses on improving malaria detection from patches segmented from microscopic images of red blood cell smears by introducing a deep convolutional neural network. Compared to the traditional methods that use tedious hand engineering feature extraction, the proposed method uses deep learning in an end-to-end arrangement that performs both feature extraction and classification directly from the raw segmented patches of the red blood smears. The dataset used in this study was taken from National Institute of Health named NIH Malaria Dataset. The evaluation metric accuracy and loss along with 5-fold cross validation was used to compare and select the best performing architecture. To maximize the performance, existing standard pre-processing techniques from the literature has also been experimented. In addition, several other complex architectures have been implemented and tested to pick the best performing model. A holdout test has also been conducted to verify how well the proposed model generalizes on unseen data. Our best model achieves an accuracy of 97.77%±0.007.**

# Introduction

Malaria, a mosquito borne life-threatening disease, causes fever, vomiting, headaches and fatigue; in severe cases it can cause coma or even death [1]. The disease can affect humans or other animals. The disease is commonly transmitted by female anopheles' mosquitoes. The mosquito bite injects the parasite into affected person's blood which then travels to the liver to mature and reproduce [2]. Malaria is caused by a single cell microorganism belonging to the genus *Plasmodium* where five of their species can infect humans. *P. falciparum* is the deadliest among these species; others include *P. vivax, P. ovale, P. knowlesi* and *P. malariae* [2].

Malaria is widespread in tropical and subtropical regions, especially in Latin America, Sub Saharan Africa and Asia [2]. The disease is considered endemic in 13 districts out of 64 in Bangladesh, putting about 14 million people at risk [3]. Among the five species, the most dominant parasite in Bangladesh is *P. falciparum* [4]. In 2016, approximately 731,000 deaths have been reported around the world due to malaria with 90% of them in Africa [5].

Malaria is commonly diagnosed by microscopic examination of blood cells using blood films [6]. Approximately, 167 million blood films had been tested for malaria during 2010 using microscopy, which was less costly and less complex than polymerase chain reaction-based diagnosis [7]. Although it is widely used, microscopic diagnosis has many drawbacks as follows. As malaria is generally associated with poverty and occurs mostly in low economic countries [8], most laboratories or diagnostic facilities are not equipped with standard testing facilities. Moreover, the diagnosis depends on the skill of the person examining the blood film and level of parasites present thereon. Additionally, the monotonicity of the examination hugely affects the quality of examination, towards the end of a batch especially if the batch has many specimens. The global shortage of pathologist [9] in general has a serious impact on health care system of developing countries and the case of malaria is no exception. Due to the lack of reliable diagnostic facilities, many Bangladeshi citizens opt for treatment overseas [10] which unfortunately is not economically feasible for majority of the people.

Current modern computer aided systems use deep learning algorithms for medical image analysis [11]. Throughout the world there is a trend to automate the diagnostic system with the help of various machine learning techniques in order to aid human specialist in making the correct diagnosis. Liang et al. have proposed a deep learning-based approach for the classification of malaria infected cells from red blood smears. Their proposed method is based on a 16-layer convolutional neural network that outperforms their transfer learning-based model using the AlexNet architecture [12] pre-trained on the CIFAR-100 dataset [13]. Dong et al. [14] have used a dataset comprising around 1000 training and testing samples only, and hence have employed transfer learning and reported the results on LeNet [15], AlexNet [16] and GoogleNet [17] architectures. Jane et. al [18] have proposed a different approach based on object detection and have used a Faster Region based Convolutional Neural Network (Faster R-CNN) that has been pre-trained on Imagenet [19] and fine-tuned on their dataset. A classic approach has been proposed by Bibin et. al [20] on deep belief network (DBN) [21] consisting of 4 hidden layers pre-trained by stacking restricted Boltzmann machines [22] using contrastive divergence method [23] for pre-training. Razzak et. al [24] have proposed an automated process that considers the tasks of both segmentation and classification of malaria parasites. Their segmentation network consists of a Deep Aware CNN [25] and the classification network employs an Extreme Learning Machine (ELM) [26] based approach.

Convolutional Neural Networks have been used in several works [27][28][29] to diagnose malaria parasites from microscopic images. Furthermore, different approaches have been proposed by Shen et. al [30] and Mohanty et. al [31] that use unsupervised machine learning approaches using stacked auto-encoders to automatically learn the features from input images of infected and uninfected cells. Mehanian et. al [27] have utilized a suite of computer vision techniques, such as, global white balance, adaptive nonlinear grayscale, to present a novel augmentation scheme which demonstrates state of the art performance for the evaluation of automated diagnosis methods.

Since medical images/datasets are usually smaller in size and thus often termed inadequate for learning, the power of transfer learning has also been leveraged in the literature. In the literature Var et al. [28] and Rajaraman et al. [29] proposed methods for computer aided diagnosis based on pre-trained convolutional neural networks as feature extractors to identify malaria parasites. Classical machine learning algorithms have also performed well in this classification task. As shown by Das et al. [32] and Park et al. [33], Bayesian learning, support vector machines, logistic regression and k-nearest neighbor algorithm perform well in this context. In addition, attempts have been made to remove the stains from the peripheral blood smear images as well as for impulse noise reduction. Mustafa et. al [34] have proposed a pre-processing step, thresholding, which is considered to be one of the most important preprocessing steps in this task. In their work, comparison among Fuzzy C-Mean algorithm [35], Wolf's method [36], Bradley's method [37] and Bernsen's method [38] has been shown. Each of these methods is experimented with malaria parasite images.

In this article, several deep neural network architectures are designed for the purpose of malaria detection in a much more improved way. We show that, using different preprocessing techniques such as standardization, normalization, stain normalization does not contribute to the model having overall performance. Instead, data augmentation techniques applied on the training set shows promising results. We show that using transfer learning, which is considered a prominent technique in computer vision, it is entirely possible to achieve certain rather good performance compared to other traditional machine learning techniques which require rigorous feature engineering and complex data pipelines as seen in literature.

Segmented patches from thin red blood cell smears have been used directly as an input, followed by deep learning in an end-to-end arrangement for the task of predicting malaria parasite. We have conducted extensive experiments using the NIH malaria dataset [29] on three different settings, namely, custom network from scratch, fine tuning on pre-trained model and CNN as feature extractor with SVM as the classifier. Furthermore, several preprocessing and post-processing methods are employed to maximize performance in an unbiased test set. To prove the robustness and invariance in different test sets, we perform 5-fold cross validation and tested the model over 5 different test sets, with each fold having 24,802 training samples and 2756 test samples. Our best model performed $0.9701\pm0.007$ for these different test sets, which ensured the generalizability of the model.

**Materials & Methods**

**NIH Malaria Dataset**

The dataset used in this work is taken from National Institute of Health [29]. It consists of segmented cells from the thin blood smear slide images from the Malaria Screener research activity. Giemsa-stained thin blood smear slides from 150 P. falciparum-infected and 50 healthy patients were collected and photographed at Chittagong Medical College Hospital, Bangladesh. The images were manually annotated by an expert slide reader at the Mahidol-Oxford Tropical Medicine Research Unit. The dataset contains 27,558 segmented cell images, with equal instances of 13,779 parasitized and 13,779 uninfected segmented red blood cell images. Positive samples contained plasmodium and negative samples contained no plasmodium but could contain other types of objects including staining artifacts/impurities. The patches of segmented red blood cells are of 3-channels (RGB) with size variation of 110-150 pixels which have been later re-sampled to 200 x 200 output dimension, a channel depth of 3 and 32-bit floating point precision (FP32) to suit the input requirements of different classification algorithms used in this work. Different pre-processing techniques are also applied to achieve faster convergence which will be discussed later in detail. Figure 1 and Figure 2 show some samples from the dataset containing uninfected and parasitized segmented red blood cells respectively.

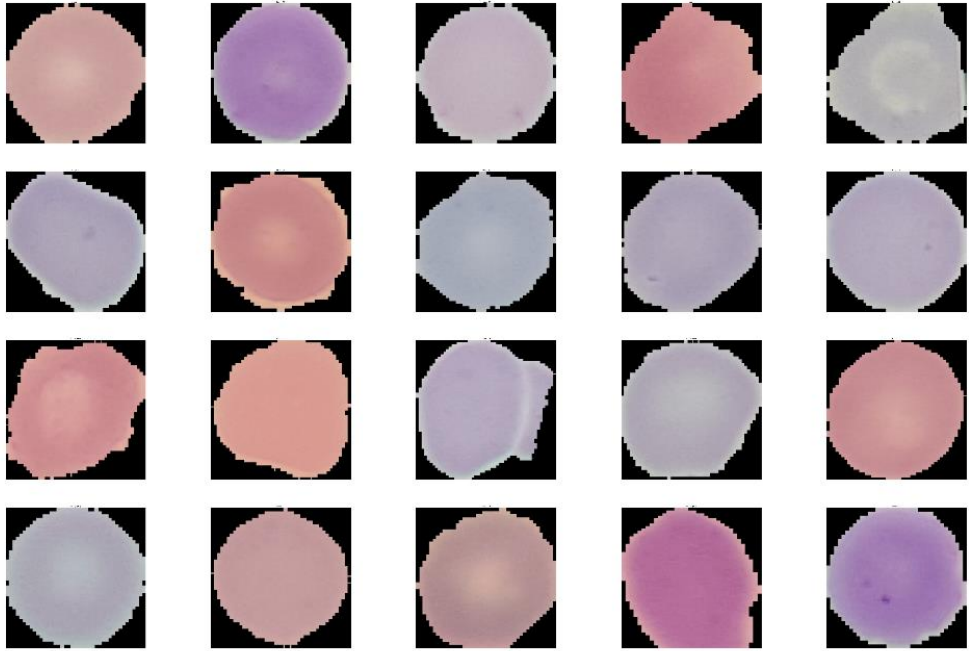

**Fig 01.** *Samples drawn from NIH Malaria dataset which are uninfected red blood cells. It is seen that the images have varying color distributions which are resulted from different stains during data acquisition.*

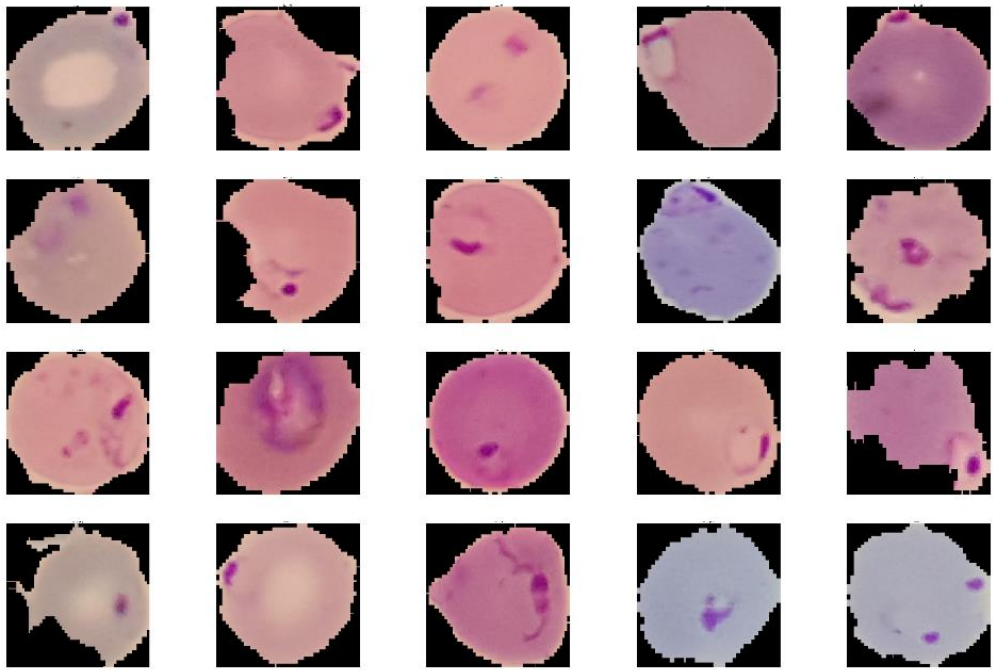

**Fig 02.** *Samples drawn from NIH Malaria dataset which are malaria infected parasite red blood cells. The images show various forms of parasite in the red blood cells.*

**Experimental Setup**

All of these experiments were performed on a machine with Windows® system with Intel® Core(TM) i7-7700 CPU @3.60GHz processor, 1 TB HDD, 64 GB RAM, a CUDA-enabled Nvidia® GTX 1060 6GB graphical processing unit (GPU), Python® 3.6.7, Keras® 2.2.4 with TensorFlow® 1.12.0 backend, and CUDA compilation tools, release 10.0, V10.0.130 dependencies for GPU acceleration.

**Evaluation Metrics**

Our proposed models were evaluated using the following standard metrics: Accuracy, Loss, Precision, Recall, F1 score, Area under curve (AUC) and Matthews correlation coefficient (MCC). In our experimental setup, we use accuracy as our optimizing metric and the others are used as satisficing metrics.

**Data Splitting**

The dataset is split into three sets, namely, training, validation and test sets having the ratio of 80:10:10. We performed both 5-fold cross validation and 5-fold holdout test to check the robustness of the network architectures under consideration. The value of k (in K-fold cross validation) is chosen to be 5 for both cases because empirically this value resulted in error rate estimates that is neither from excessively high bias nor from very high variance. At this point a brief discussion is in order. It might seem counter-intuitive that in our experiments, 5-fold cross validation performed better. However, we argue that in the context of medical images this is not surprising as discussed below. Medical images are comparatively more diverse and subjective and same case is observed with our dataset as well. Therefore, it is almost always expected that when deploying a medical imaging algorithm to a real-world scenario, we are likely to receive images drastically different from those in the training set. Now, using 5-fold CV instead of 10-fold CV or LOOCV provides us a broader test set and lesser amount of training data in this context which is somewhat mimicking the real-world scenario.

As mentioned earlier, to evaluate the robustness of the network architecture, a 5-fold cross validation has been done. Here the training dataset is randomly partitioned into 5 equal sized subsets; of the 5 subsets, a single subset is retained as the validation data for testing the model, and the remaining 4 subsets are used as training data. The cross-validation process is then repeated 5 times, with each of the 5 subsets used exactly once as the validation data. The 5 results can then be averaged to produce a single estimation. The standard deviation is then used for the main evaluation of the model.

To further validate if the network architectures generalize on different datasets, a 5-fold holdout test has been conducted like that of cross validation. Here, the whole NIH Malaria Dataset is split in the ratio of 80:10:10 for training, validation and testing set. This was randomly sampled from the dataset in each of the 5 folds.

**Data Preprocessing**

In this section, several pre-processing techniques are discussed in detail, which have been experimented with, and the outcomes are reported.

**Stain Normalization**

For a technician to examine malaria parasite from blood film using microscope, the films are prepared by staining them with various chemical stains, such as giemsa stain, wright stains [39][40]. Color variations exist due to the use of different chemical stains, scanners, stain concentration, staining procedure. This causes the network to learn an even more complex function which can result in greater error margin. A standard solution to this problem is standardization. This requires the normalization of blood films. Khan et. al have proposed a method to normalize stains on histopathology images [41]. Subsequently, Ciompi et. al have shown how stain normalization can improve classification of colorectal tissues in colorectal cancer [42]. Stain normalization is experimented in this work on the patches of red blood cell smears.

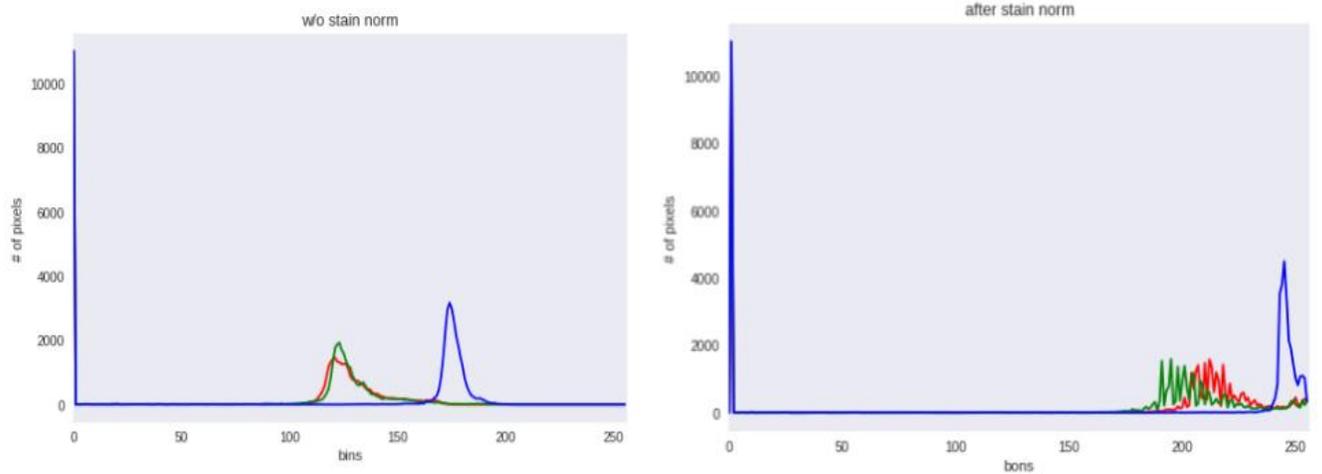

**Fig 03.** *Intensity distribution of an isolated cell before and after stain normalization. Each graph represents corresponding color channels (Red, Blue & Green). The y-axis is the number of pixels and x-axis is the range of pixel values (0-255). Stain normalization results to a narrow domain without changing the semantic meaning of the image.*

Figure 3 shows a random red blood cell patch before and after stain normalization. The histogram represents the RGB color space in which the image lies in the high dimensional feature space. The y-axis is the number of pixels and the x-axis is the range of pixel values (0-255) which is divided into 16 bins to represent the histogram. It is evident from the graph that after stain normalization, the image is transformed to a narrower range in the high dimensional feature space while preserving the semantic meaning shown in Figure 4.

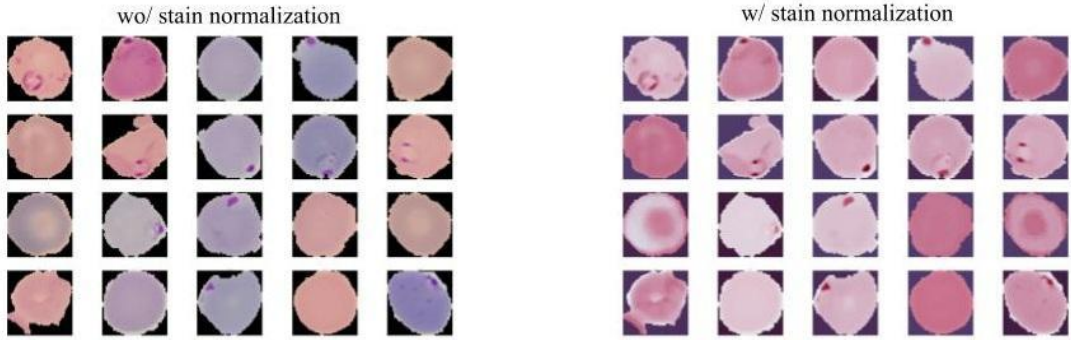

**Fig 04.** *Before and after stain normalization applied to red blood cell (RBC) patches. All RBCs have been transformed to have less color variations. This reduces the stain variation from training images while preserving the semantic meaning of images.*

From the figure it is evident that the semantic meaning is preserved after stain normalization is applied to the red blood cell patches. It is also seen that the stain is removed after this transformation and the data points have less color variation.

**Rescaling (Min-Max Normalization)**

In order to achieve a faster convergence, the patches of images were rescaled to map the range of features to 0 to 1. Since the red blood cell patches are 8-bit RGB color images, the maximum pixel value is 255 and the minimum is 0. This rescaling is achieved by the following equation:

$$X = \frac{X - Xmin}{Xmax - Xmin} = \frac{X}{255} \quad (1)$$

**Standardization**

Another pre-processing technique named feature standardization has been experimented with. Here the values of each feature in the data is rescaled to have mean 0 and standard deviation 1. Both the training and validation patches were standardized by using Equation 2:

$$X = \frac{X - \mu}{\sigma} \qquad (2)$$

Here standard deviation is denoted by *σ* and mean is denoted by *μ*. Both mean and standard deviation are calculated from the corresponding training set.

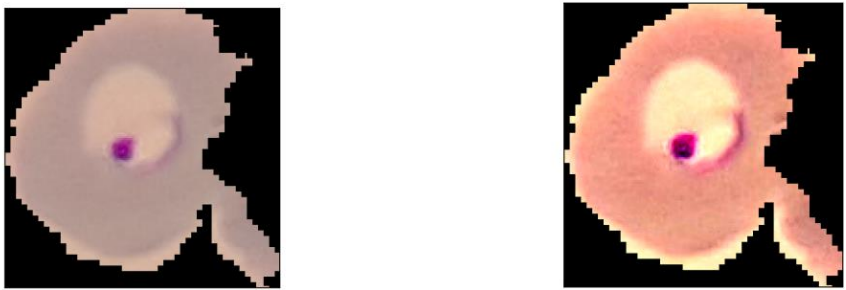

**Fig 05.** *Before and after standardization of RBC patch. Images have been rescaled to have mean 0 and standard deviation of 1. Earlier, each RBC image had pixel values between 0 to 255.*

Figure 5 shows an instance of a segmented RBC patch before and after the standardization is applied.

**Data Augmentation**

Data augmentation refers to enhancing the base data to increase data points. It has been used in various medical dataset to improve classification performance [43][44]. We have used data augmentation to improve performance during the hold out test. The dataset consists of equal instances of their respective classes; thus, it is balanced, this contribute towards solving the problem of overfitting during training. Although the training set had much larger number of training samples, they were not representative of the test set. Certain image augmentations have been applied to the training data while preserving the semantic meaning of the segmented patch as follows.

Data augmentation applied in our case includes horizontal and vertical flips, gaussian blur, rotation, horizontal and vertical shifting, darkening and lightening, ZCA whitening and feature wise standardization. Even though feature standardization is a preprocessing technique as discussed in previous section, we use this technique in our augmentation pipeline. It is noteworthy to mention that; the feature standardization has not been used as a preprocessing technique at the start of training. Change of color space and Gaussian Blur have been performed with a probability of 0.5 and the rest have been performed in randomized order on each image as summarized in Table 1.

**Table 01.** Augmentation types and parameters

| Augmentation Type | Parameters |
|---|---|
| Contrast Normalization | (0.5, 1.5) increase/decrease in range |
| Crop | (0, 0.2) 0-20% of their height/width |
| Rotate | (-25, 25) degrees in range |
| Translate Percent | x: (-0.2, 0.2), y: (-0.2, 0.2) move by this range |
| Shear | (-25, 25) |
| Color Space | HSV to RGB |
| Gaussian Noise | (0.0, 0.05*255) Sample noise once per pixel |

After augmentation, the initial training set of 27,558 images extends to 137,940. Figure 2 shows an example of an original image with different types of augmentations.

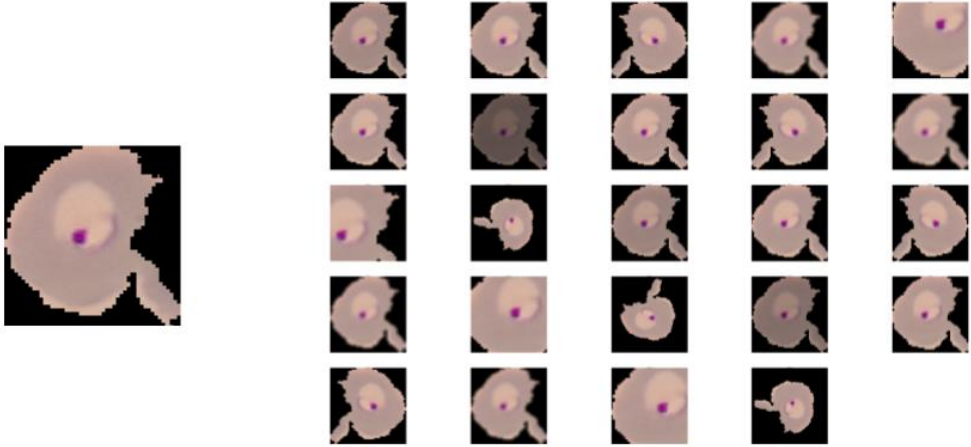

**Fig 06.** *Augmented Images of a single malaria blood sample to increase data points. Each image has been flipped, rorated, translated, blurred, cropped etc. to have better representation of test set.*

**The Network Architecture**

As has been alluded to in an earlier section, we have employed three different experimental settings. We have built a custom network from scratch (referred to as Custom henceforth). We also have leveraged transfer learning by fine tuning on a pre-trained model, namely, VGG16 [45] (referred to as TL-VGG16 henceforth). Finally, we have explored the capability of CNN as feature extractor and have employed SVM as the classifier using the extracted deep features (referred to as CNNEx-SVM henceforth). Followed by, we will describe each of these elaborately.

**Custom**

A custom architecture has been designed consisting a total of 19 layers, with 8 convolution layers, 4 max pool layers, 3 dense layers, one flattens layer, two layers with 50% dropout (to reduce overfitting) and one fully connected layer as shown in Figure 7. Each of these convolutional blocks is input to an activation function. Here as well Rectified Linear Units (ReLU) is used as the activation function as follows:

$$a = max\ (0,\ x) \qquad (3)$$

Here, a is the output activation for a given input *x*.

In the convolutional operation, a filter of size of 3 x 3 with padding set to same is used to convolve over the input volume. A pool size of 2 x 2 is used in all the max pooling layers. The output from the final max pooling layer is input to a flatten layer with 36864 output neurons. This is followed by two blocks of a dense layer of 256 output neurons followed by a dropout layer which randomly discards 50% of the input neurons. In the final classification stage, the output of the dense layer is input to another activation function. Here softmax activation has been used in order to get the output probabilities for the binary classification problem. For our binary classification problem, the loss function binary cross entropy is used which calculates the error between the ground truth and predicted output probabilities. The adaptive learning rate, named Adadelta optimizer [46], is used to optimize the weights and biases of the network via backpropagation. Initially the weight is initialized randomly as in [47] and the biases are set to 0. The activations are computed for a batch size of 64 samples. This is continued for 30 epochs.

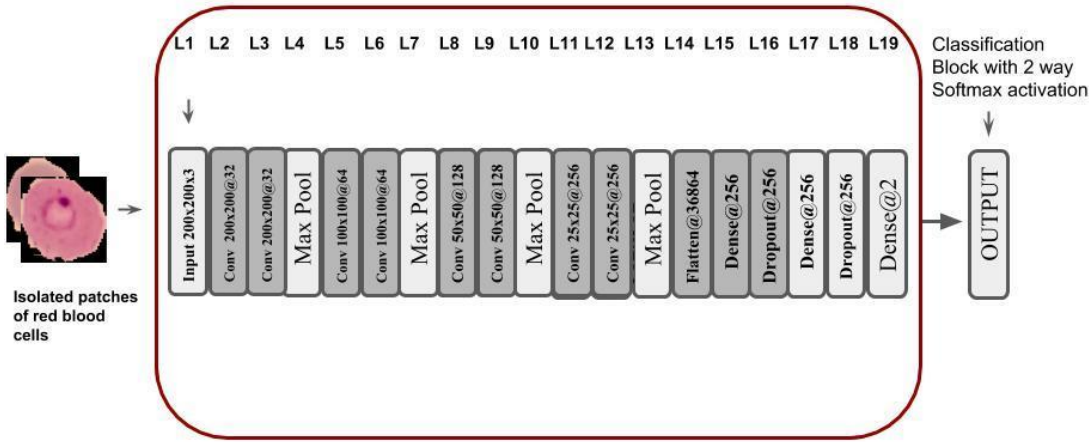

**Fig 7.** *19 layers custom architecture of Neural network with 2-way softmax activation. The input dimensions are 200 x 200 and ReLu is used as activation for convolution layers.*

Table 2 shows a summary of the model settings along with the hyperparameter values.

**Table 2.** Summary of the model settings along with the hyperparameter values

| Parameter Name | Type/ Value |
|---|---|
| Epochs | 30 |
| Batch Size | 64 |
| Optimizer, Learning Rate | Adadelta |
| Error Function | Binary Cross Entropy |
| Input Size | 200 x 200 |
| Pooling | 2 x 2 |

**TL-VGG16: Transfer learning (VGG16 as a baseline):**

To leverage transfer learning, the VGG16 [45] architecture along with its weights that were pre-trained on ImageNet has been used in this work. This architecture consists of 16 convolutional layers. Each of these convolutional blocks is input to an activation function. Here as well Rectified Linear Units (ReLU) is used as the activation function as well. We have updated the network as follows. From the VGG16 architecture, originally present fully connected layers pre-trained on ImageNet with a 1000-way softmax activation are discarded. This is followed by adding a dense layer with 1024 output neurons which is input to a dropout layer [48] that drops neuron connections with a probability of 50%. The output from the dropout layer is followed by a single fully connected layer with 2-way softmax activation to get the probability results for the binary classification problem. The figure below shows the VGG16 architecture after modification with all the blocks.

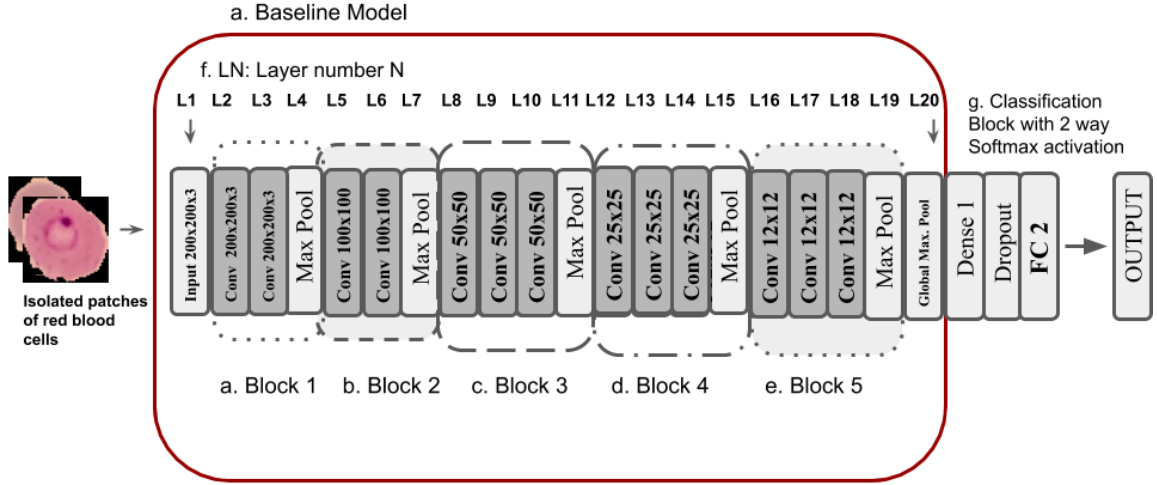

**Fig 8.** *VGG-16 architecture as a baseline with 2-way softmax activation. The input dimensions are 200 x 200 and ReLU is used as activation for convolution layers.*

The adaptive learning rate method, named Adadelta [46], is used to optimize the weights and biases of the network with a starting learning rate of $10^{-2}$. Adadelta is a more robust extension of other optimizers such as Adagrad, that adapts learning rates based on a moving window of gradient updates, rather than collecting all past gradients. The cross-entropy error [49] between predicted and ground truth has been used as a loss function and for the binary classification problem. It is calculated as shown in Equation 4 and the weights are updated using mini batches of 128 training samples at every iteration.

$$\text{Cross-Entropy} = -\,(y\,log(p) + (1-y)\,log(1-p)) \qquad (4)$$

Here *log* is natural logarithm, y is the binary indicator (0 or 1) indicating whether class label *c* is the correct classification for observation *o* and *p* is the predicted probability that observation *o* is of class *c*.

At the beginning of the training loop, the weights are initialized as proposed in [46] and the biases are initialized to zero.

Table 3: Layer tweaking in VGG16

| Layers Frozen | Training Loss | Training Acc | Validation Loss | Validation Acc | Test Loss | Test Acc |
|---|---|---|---|---|---|---|
| L1-L20 | 0.1859 | 93.58 | 0.1856 | 93.94 | 0.195 | 93.57 |
| None | 0.0085 | 99.86 | 0.1582 | **96.70** | 0.1741 | 95.60 |
| L1-L8 | 0.0085 | 99.86 | 0.1576 | 96.66 | 0.1725 | 95.71 |
| L1-L16 | **0.0019** | **99.95** | 0.2092 | 96.66 | 0.1620 | **97.64** |
| L1-L14 | 0.0448 | 98.53 | **0.1013** | 96.66 | 0.1229 | 95.64 |

The architecture was fine-tuned by training only the last four layers (L17 to L20 of Figure 8) of the VGG16 architecture. Table 3 portrays the results of a series of ablation experiments conducted by freezing intermediate layers [50] of the network as mentioned in Table 3. It is evident from the ablation experiments that the model with L1-L16 layers frozen outperforms the other settings. This final architecture specification is summarized in Table 4.

**Table 4:** VGG16 based model configuration and Hyper parameter setting

| Parameter Name | Type/ Value |
|---|---|
| Epochs | 50 |
| Batch Size | 64 |
| Optimizer, Learning Rate | Adadelta, 0.01 |
| Error Function | Binary cross entropy |
| Input Size | 200 x 200 |
| Pooling | Average |
| Pre-trained Weights | ImageNet |
| Layers Frozen | L1 to L16 (Refer to Figure 08) |

Training was continued for 50 epochs in this configuration and only the best weights for the highest accuracy on the validation set was saved as model parameters.

**CNNEx-SVM: CNN features applied to SVM**

Support vector machine is a supervised machine learning algorithm used for classification and regression problems. Besides linear classifications, Support Vector Machine can also perform nonlinear classifications by implicitly mapping their inputs into high-dimensional feature spaces [51]. The extracted features from convolutional neural networks seem to work well if an SVM is used to do the classification instead of softmax layer [52] [53] [54].

Features from TL-VGG16 are extracted after training the network with malaria dataset. The extracted features are then fitted to a Support Vector Machine. In classification using SVM, radial basis functions have been used with gamma being 0.1 for mapping the data in higher dimensional space. SVM with CNN features resulted in accuracy of 0.93 with standard deviation of 0.007. This has also been done with the features extracted from the custom network, which resulted in lower accuracy (91.13%) than by features computed by VGG16. Therefore, it can be said that the custom network cannot extract deep features compared to VGG16 pretrained on ImageNet.

**Ensemble**

To maximize the performance, the ensemble learning method has also been considered in this work. Ensemble learning combines the predictions from multiple (weaker) models to reduce the variance of predictions as well as the generalization error. Ensemble of neural networks are known to be much more robust and accurate than individual neural networks. Similar work also has been found in the literature [ e.g., 55, 56, 57, 58]. Figure 9 depicts a schematic diagram of our ensemble-based approach. The output predictions from each of the three networks are concatenated and a weighted average is taken to give the final prediction. Table 5 shows the result of applying ensemble of the three proposed architecture.

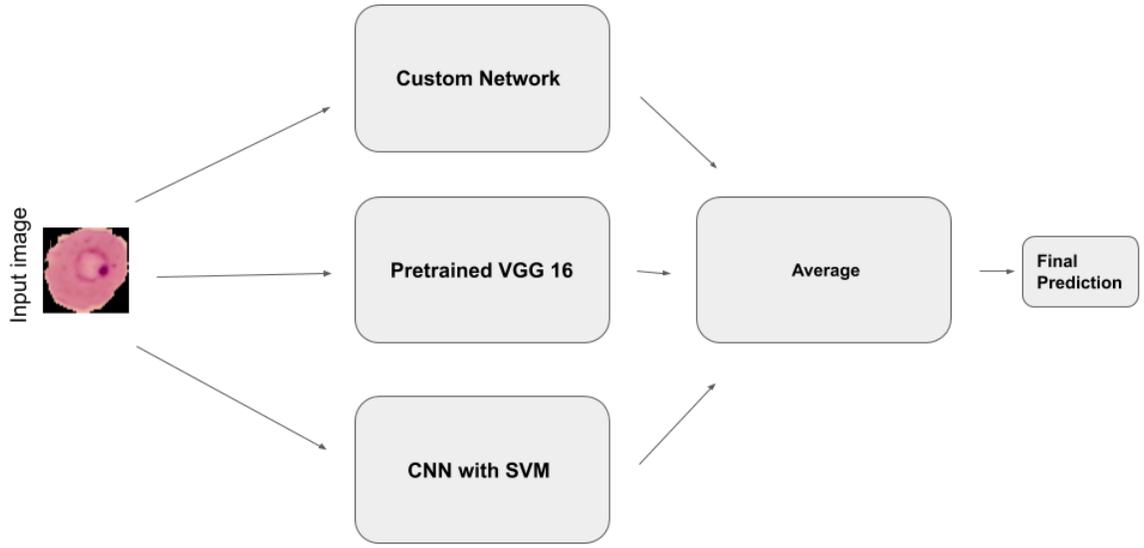

**Fig 9.** *Ensemble of three models and their combined predictions. The three different deep learning models used in this experiment (Custom CNN, TL-VGG16 & CNN-Ex SVM) are combined to produce a better prediction for test set.*

**Table 05.** Comparisons of Ensemble of CNNs with different models

|  | TL-VGG16 | Custom | CNNEx-SVM | Ensemble |
|---|---|---|---|---|
| **Test Accuracy** | 0.9764 | 0.9597 | 0.9477 | 0.9778 |

**Test Time Augmentation**

Test Time Augmentation is a way of improving the performance of deep learning models [59]. The goal of test time augmentation is to perform some random augmentations of the test images. Instead of predicting on only one original image, the model predicts on several randomly augmented images and result is based on some average of those predictions. We have performed several different augmentations, similar to what has been discussed in Table 1, during inference and the model is exposed to slightly modified versions of the test image and a weighted average consisting of equal weights for each augmentation is calculated from the individual predictions to give a final prediction.

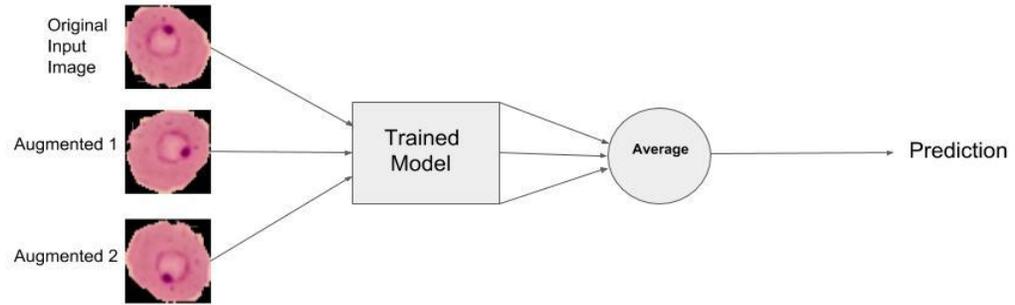

**Fig 10.** *Test Time Augmentation used as post processing. Images have been augmented several times and their weighted average predictions are considered.*

**Table 06.** Performance analysis using Test Time Augmentation using different models

|  | **TL-VGG16** | **Custom** |
|---|---|---|
| **Test Accuracy After TTA** | 0.9789 | 0.9645 |

**False Case Analysis**

In this section, we show an error analysis performed on our best performing model. The predicted values are compared to the ground truth values by showing the false negatives and false positives (please see Figure 11). From Figure 11 we can see that our best performing model misclassified a total of 63 images when evaluated on the test set which consists of 2756 samples. It is noteworthy, that the same preprocessing techniques are applied during inference that was used while training.

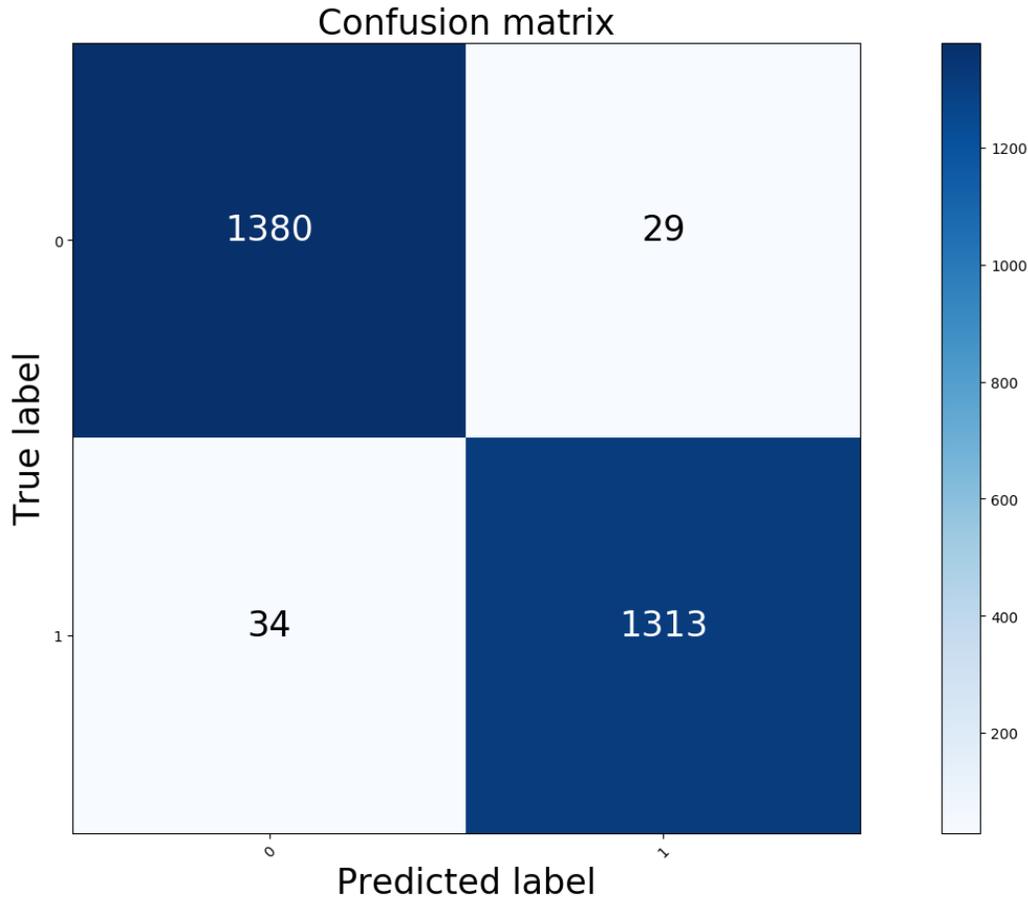

**Fig 11**: *Confusion matrix of TL-VGG16 model. The numbers represent true positive, false positive, false negative and true negatives respectively.*

From the confusion matrix it is evident that the number of false negatives and false positives are 34 and 29 respectively. Figure 13 presents the images which the model misclassified. From Figure 12, it is observed that in some cases the images which are labelled as uninfected are in fact very similar to parasite samples in the test set, which strongly suggests the possibility of mislabeling by the expert while labeling the dataset. To confirm this claim, the false positives from TL-VGG16 have been evaluated by a pathologist. Following evaluation, it has been concluded that 38% of the blood cells from the false positives are in fact mislabeled, showing clear formation of malaria parasites while labelled as uninfected. We can also say that a similar error rate can be found for the false negatives. Due to budget constraints and large number of blood cell images, we were unable to perform a thorough data validation scheme conducted by a pathologist over the entire NIH Malaria dataset. This is a limitation in our series of experiments demonstrated in this work.

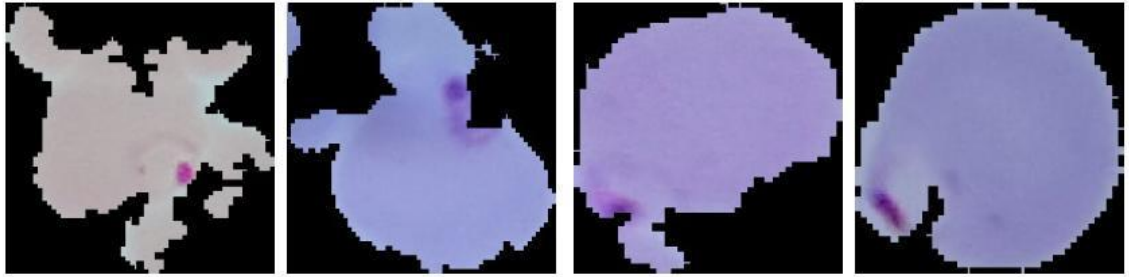

**Fig 12:** *Example of mislabeled parasite infected blood cells confirmed by pathologist. Images shown here have a ground truth of normal but are evaluated as parasite by a pathologist.*

In the cases of false positives, it is seen that the model miss-classifies the red blood cells which are in an early stage of malaria - consisting of only a few parasites. Furthermore, we show from error analysis that the mislabeled data also has a negative impact in the overall performance of the system. Perhaps this problem can be mitigated by treating this perception task as a multi class classification problem where a severity-based scoring may be introduced. However, this is beyond the scope of the current work and will be briefly discussed as a future work in a forthcoming section.

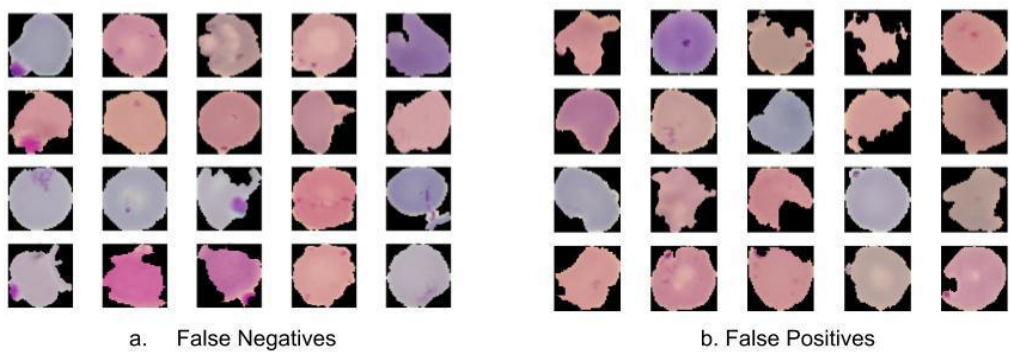

**Fig 13**: *Samples depict the predictions using the TL-VGG16 architecture. a) Images predicted by the network which are normal but predicted as parasite b) Images predicted by the network which are parasite but in fact were classified as normal by the architecture.*

**Individual Patient Diagnosis:**

We also test our best performing model, named TL-VGG16, for individual patient diagnosis. We preprocess the metadata taken from the NIH Malaria Dataset which list each patient and their corresponding red blood cell images, both parasitized and uninfected. For each patient, we run their corresponding images to our cell level classifier and report the diagnosis as such: if any datapoint is classified as parasite cell, the patient is assigned a class value of parasite; meaning the patient is exposed to malaria. In our experimental setup, our best performing model, TL-VGG16, achieves an accuracy of 93% with and Area Under Curve (AUC) of 86.5%. Notably, the degradation in performance for individual patient diagnosis can be attributed to the staining variations between patients. We have shown in the previous section that it is harder for the classifier to learn the different stains, resulting in misclassifications.

**Results**

Table 7. Different network architecture results on each of the metrics

| Model | Test Accuracy | Precision | Recall | F1 score | MCC | AUC |
|---|---|---|---|---|---|---|
| Custom | 96.29 | 0.9804 | 0.9234 | 0.9495 | 0.9051 | 0.9116 |
| TL-VGG16 | **97.77** | **0.9719** | **0.9720** | **0.9709** | **0.9442** | **0.9938** |
| State-of-the-art Customized [29] | 94.00 | 0.951 | 0.931 | 0.941 | 0.880 | 0.979 |
| State-of-the-art ResNet-50 [29] | 95.70 | 0.969 | 0.945 | 0.957 | 0.912 | 0.990 |
| CNNEx-SVM | 94.77 | 0.9213 | 0.9515 | 0.9501 | 0.8925 | 0.9101 |

We have evaluated our network architectures on different metrics, namely, Accuracy, Loss, Precision, Recall, F1 score, Area under curve (AUC) and Matthews correlation coefficient (MCC) and the results are tabulated in Table 7. From Table 7, it is evident that TL-VGG16 outperforms the other architectures as shown with a test accuracy of 97.77%. On the other hand, Custom model performed with 96.27% accuracy and 94.77% for CNNEx-SVM. The test set scored 94% and 95.70% for current state of the art Custom and ResNet-50 model.

The experimental results on the different models with stain normalization is tabulated below in Table 8.

Table 8. Performance analysis using stain normalization over different models

| Model | Accuracy w/ SN | Accuracy w/o SN |
|---|---|---|
| Custom | 0.9499 | 0.9629 |
| TL-VGG-16 | 0.9694 | 0.9727 |
| CNNEx-SVM | 0.9241 | 0.9477 |

Table 8 shows the impact of the stain normalization preprocessing technique (applied on the NIH Malaria Dataset) on our proposed models. Stain normalization does not provide any improvement for any of the models.

Table 9. Performance of different Normalization techniques

| Model | Standardization | Mean Normalization |
|---|---|---|
| Custom | 0.9593 | 0.9560 |
| TL-VGG16 | 0.9702 | 0.9738 |
| CNNEx-SVM | 0.9241 | 0.9328 |

In Table 9, we compare the effectiveness of standardization and mean normalization. It can also be seen here that after employing standardization and mean normalization, we do not get good results. It can be said that both preprocessing techniques do not add impact in our setting. The ranges of the data distributions of feature values likely be not different for each feature, hence standardizing/normalizing impact on these networks. Figure 14 shows the training accuracy and loss for the custom model. The model is trained for 30 epochs and the model converges in just 10 epochs with a validation accuracy of 95%. As the model progresses to train, there can be seen signs of overfitting and the loss starts to increase.

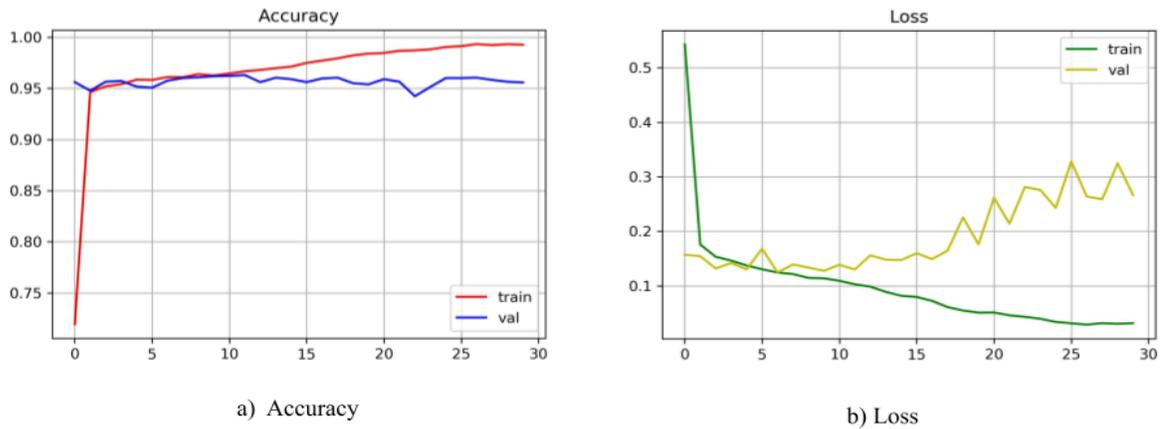

**Fig 14**. *Training and validation logs of the custom model after running for 30 epochs. a) Accuracy plotted in every epoch throughout the training regimen. The network achieves a training accuracy of 99% and validation accuracy of 94.5%. b) Loss plotted in every epoch showing a maximum validation loss of 0.3.*

Since we leverage the power of transfer learning, the VGG16 architecture that we use as a baseline is trained essentially for 100 epochs. This is mainly because the architecture contains a greater number of parameters compared to the custom model and it can be trained effectively longer to extract deep features. From Figure 15, the model converges at about 80 epochs of training and achieves a validation accuracy of 97.58% and a validation loss of 0.1.

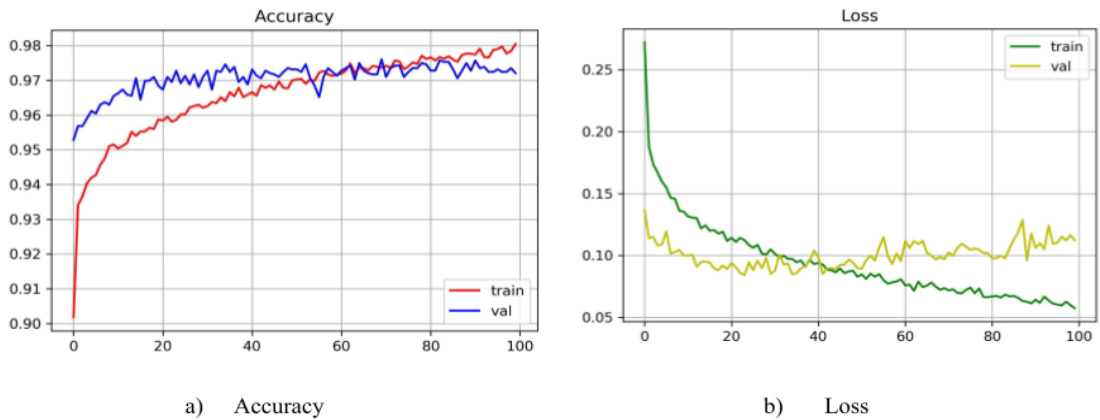

**Fig 15.** *Training and validation logs of the TL-VGG16 model after running for 100 epochs. a) Accuracy plotted in every epoch throughout the training regimen. The network achieves a training accuracy of 98% and validation accuracy of 97.7%. b) Loss plotted in every epoch showing a maximum validation loss of 0.12.*

**Discussion**

- A. We have experimented with several preprocessing and post-processing techniques to analyze their effectiveness on malaria dataset. Performing normalization and standardization did not improve the accuracy of the models, although models converged faster when these processing techniques were applied. These results suggest that the features do not have to be in the same scale for the models to be trained.
- B. Stain normalization, another pre-processing technique, also did not increase the performance on malaria dataset. While this may seem counter-intuitive, we believe this is due to the fact that stain normalization technique applied here was originally developed for H&E stains; the samples from the NIH malaria dataset on the contrary does not have enough Hematoxylin and hence the technique was not useful here. Furthermore, in order to pick the stains, higher luminous pixels are considered background and filtered by the algorithm. So, the final image may not look exactly like the original one, thereby changing the semantic meaning of the image. Since the dataset have enough instances of each of the stains, the models worked better on dataset without stain augmentation.

C. General data augmentation techniques performed significantly better on the NIH Malaria dataset. Partial obstruction, drop-out was not performed in augmentation technique as these techniques might have removed parasite information's from RBCs.
D. Post processing techniques such as ensemble and test time augmentation both have improved model performance. In order to check whether these models are biased to a specific test, 5 different independent test sets were evaluated on 5 different models. Hence, our best performing model do not have larger variance from the validation sets.
E. In Table 10, we show results after performing a 5-fold cross validation and a holdout test of multiple instances over the NIH Malaria Dataset. Evidently, in 5 different test sets our best model TL-VGG16 does not have much variation thus ensuring robustness and generality.

**Table 10**. Performance evaluation

| Model | Holdout Test | Cross Validation |
|---|---|---|
| TL-VGG16 | $0.9701 \pm 0.007$ | $0.9700 \pm 0.005$ |
| State-of-the-art [29] as reported on paper | - | $0.957 \pm 0.007$ |

**Conclusions**

In this work, we conduct a series of experiments based on end-to-end deep learning to improve malaria classification from segmented red blood cell smears. We show that, using different preprocessing techniques such as standardization, normalization, stain normalization does not contribute to the model having overall performance. Instead, data augmentation techniques applied on the training set shows promising results. In addition, we demonstrate different network architectures namely: Custom network architecture, Fine tuning on pre-trained models and extracting features from a convolutional network (CNN) followed by a support vector machine classifier (SVM). From our proposed models, TL-VGG16 achieves an accuracy of 97.77% on the hold out test which outperforms all the other models. We show that using transfer learning, which is considered a prominent technique in computer vision, it is entirely possible to achieve certain rather good performance compared to other traditional machine learning techniques which require rigorous feature engineering and complex data pipelines as seen in literature. In future works, the network architecture is to be further analyzed to understand why the results occurred.

**Acknowledgement**

M.R.C. Mahdy acknowledges the support of North South University internal research grant 2018-19 along with the support of Dhaka Central International Medical College and Hospital, 2/1 Ring Road, Shyamoli, Mohammadpur, Dhaka-1207.